\documentclass[onecolumn]{elsart}
\usepackage{graphicx,amssymb,amsmath}
\usepackage[square,comma,sort&compress]{natbib}
\usepackage[dvips,usenames]{color}
\usepackage{fancyvrb}

\newcommand{\eg}{{\it e.\ g.}}
\newcommand{\ie}{{\it i.\ e.}}
\journal{Computer Physics Communications}

\begin{document}

\begin{frontmatter}
\title{The Truncated Polynomial Expansion
Monte Carlo Method for Fermion Systems Coupled to Classical Fields:
A Model Independent Implementation}
\author[ornl2]{G. Alvarez},
\author[nhmfl]{C. \c{S}en},
\author[aoyama]{N. Furukawa},
\author[riken]{Y. Motome}
and
\author[ornl,utk]{E. Dagotto\thanksref{nhmfl2}}
\thanks[nhmfl2]{On leave, from  Dept. of Physics,
Florida State University, Tallahassee, FL 32306}

\address[ornl2]{Computer Science \& Mathematics 
Division, Oak Ridge National Laboratory, \mbox{Oak Ridge, TN 37831}}
\address[nhmfl]{National High Magnetic Field Lab and Department of Physics, Florida State
University, Tallahassee, FL 32310}
\address[aoyama]{Department of Physics, Aoyama Gakuin University, Fuchinobe 5-10-1,
Sagamihara, Kanagawa 229-8558}
\address[riken]{RIKEN (The Institute of Physical and Chemical Research), 2-1 Hirosawa,
Wako, Saitama 351-0198}
\address[ornl]{Condensed Matter Sciences Division, Oak Ridge National Laboratory, Oak
Ridge, Tennessee 37831}
\address[utk]{Department of Physics and Astronomy, The University of Tennessee, Knoxville,
Tennessee 37996}

\begin{abstract}
A software library is presented for the polynomial expansion method (PEM) of the density of states (DOS) 
introduced in Ref~\cite{re:motome99,re:furukawa01}. The library provides all necessary functions for the 
use of the PEM and its truncated version (TPEM) in a model independent way. The PEM/TPEM
replaces the exact diagonalization of the one electron sector in models for fermions
coupled to classical fields. The computational
cost of the algorithm is $O(N)$ -- with $N$ the number of lattice sites --
 for the TPEM\cite{re:furukawa03} which should be 
contrasted with the
computational cost of the diagonalization technique that scales as
$O(N^4)$. The method is applied for the first time to a double exchange model with
finite Hund coupling and also to diluted spin-fermion models. 
\end{abstract}
 
\end{frontmatter}

\section*{PROGRAM SUMMARY}
{\bf (Computer program available on request)}\\
{\bf Title of library:} TPEM\\
{\bf Distribution Format:} tar gzip file\\
{\bf Operating System:} Linux, UNIX\\
{\bf Number of Files:} 4 plus 1 test program\\
{\bf Keywords: Moment expansion, Monte Carlo method, Correlated electrons} \\
{\bf Programming language used:} C\\
{\bf Computer:} PC\\
{\bf Typical running time:} The test program takes a few seconds to run.\\
{\bf External routines: } The LaPack library can be used by the test program and the GSL library
 can be used by TPEM though they are not essential.\\
{\bf Nature of the physical problem:} The study of correlated electrons coupled to classical fields
appears in the treatment of many materials of much current interest in condensed matter
theory, \eg\ manganites, diluted magnetic semiconductors and high temperature superconductors 
among others.\\
{\bf Method of solution:} Typically an
exact diagonalization of the electronic sector is performed in this type of models
 for each
configuration of classical fields, which are integrated using a classical Monte Carlo
algorithm. A polynomial expansion of the density of states is able to replace the exact
diagonalization, decreasing the computational complexity of the problem from $O(N^4)$ to
$O(N)$ and allowing for the study of  larger lattices and more complex and realistic
systems.\\
{\bf Key References:} \cite{re:motome99,re:furukawa03,re:dagotto02,re:alvarez02}.\\

\section{Introduction}
The problem of fermions coupled to classical fields appears in many contexts in
condensed matter physics. In this kind of problems the fermionic operators 
appear in the Hamiltonian involving only quadratic terms.
They can be solved \cite{re:dagotto02} by diagonalizing the fermions exactly
in the one-electron sector at finite temperature for a given configuration of classical
fields. The classical fields are integrated by means of a classical Monte Carlo algorithm.
The procedure, apart from being exact within the error bars, preserves the lattice throughout 
the calculation making it possible to study 
the spatial dependence of the observables.
This is a crucial issue to understand inhomogeneities and has been successfully applied to
the study of many materials \cite{re:alvarez04}. 
In the case of manganites such models have been used to understand
the phase diagram of these materials as 
well as the colossal magnetoresistance effect \cite{re:dagotto02},
 \ie\ the colossal
response of the system to magnetic fields, that could have 
many important applications. In this case, the classical field is the local $t_{2g}$ spin. 
Diluted magnetic semiconductors have also been studied in a similar way \cite{re:alvarez02}. 
The inhomogeneities that appear there in
the form of ferromagnetic clusters, could only be accessed by the use of this method
\cite{re:alvarez03}. In
addition, a model for high temperature superconductors 
has been presented \cite{re:alvarez04b} to study the competition between d-wave superconductivity and
antiferromagnetism that seems to explain interesting properties of these complex
materials.\\
Despite all these many advantages of the method, it is still very costly in terms of
computational effort. Indeed, the method scales as order $O(N^4)$ and the largest lattices
that can be accessed in a practical way contain no more than $6^3$ sites or its equivalent in lower
dimensions. This imposes limitations on the kind of physical systems that can be studied,
for example, the Mn spin concentration in diluted semiconductors 
has to be high enough, the study of many band
systems becomes difficult, etc.\\
Trying to solve some of these problems, 
two of the authors (N.F. and Y.M.) proposed \cite{re:motome99,re:furukawa01} in 2001 a procedure that replaces the exact diagonalization of
the one-electron sector by a series expansion of the density of states in terms of
Chebyshev polynomials. The method takes advantage of the sparseness of the Hamiltonian
matrix (which is the case in virtually all systems of physical interest) to perform the
matrix-vector products that appear in the calculation of the terms or moments of the
expansion. In what follows, this method will be referred to as the polynomial expansion
method or PEM. In 2003  an improvement of the PEM was proposed, \cite{re:furukawa03} based on
two controllable approximations that, as will be seen, do not  diminish in any way the
quality of the results. The first of these approximations is the 
truncation of each matrix-vector multiplication, including only products that are larger
than a certain threshold.
The second one is the truncation of the difference in Boltzmann probability weight or
 action between two very similar configurations of classical fields. This difference
 appears in the Monte Carlo procedure with enormous frequency and so its truncation turns out to be 
 very effective. This new truncated PEM will be referred to as TPEM.\\
In this paper we present a C library that implements both the PEM and TPEM. The library is model
independent and basically takes as input the Hamiltonian matrix of practically any model of fermions
coupled to classical fields. To our knowledge no such library is presently available but its usefulness is
evident: the TPEM can be easily separated from other details of the main program(s) and users do not have
to be concerned with the technicalities of the method. In this sense the library presented here 
places the TPEM at the same level of the exact diagonalization.\\
Another algorithm for the study of spin-fermion models on large lattices is the ``Hybrid Monte Carlo Algorithm",
that has been applied to the double-exchange model with infinite coupling \cite{re:alonso01}.
In this method the model is formulated in the path integral representation, introducing imaginary time, 
and a Hybrid Monte Carlo (HMC) procedure
is used
to evolve the system. The TPEM seems to work best than the HMC at low temperature where the HMC presents increasing 
computational cost due to the time discretization.
Furthermore, the HMC is applicable when the
bands  are connected and  do not extend over a wide range of energies, as is the case
 of finite coupling systems. The TPEM also allows for easy parallelization, improving the performance even more.\\
The paper is divided as follows. Section~\ref{sec:overview} explains the  theory underlying the TPEM.
 In Section~\ref{sec:implementation} the implementation details and the 
functions provided by the library
are described. Section~\ref{sec:simpleexamples} shows some simple examples on how to use the library.
Finally, in Section~\ref{sec:applications}, the TPEM is applied to a model for manganites with finite
coupling and also to diluted spin-fermion systems.\\
It is important to emphasize that the 
application of the TPEM to finite Hund coupling and diluted systems is 
 novel and shows that the method is suitable to study both systems with disconnected bands and
systems with impurity bands.

\section{Theoretical Overview}\label{sec:overview}

The analysis starts with a model defined by a certain Hamiltonian,
$\hat{H}=\sum_{ij\alpha\beta}c^\dagger_{i\alpha} H_{i\alpha,j\beta}(\phi)c_{j\beta}$, 
where the indices $i$ and $j$ 
denote a spatial index and $\alpha$ and $\beta$ some internal degree(s) of freedom, \eg\ 
spin or orbital.
The Hamiltonian matrix depends on the configuration of one or more classical fields,
represented 
by $\phi$. Although no explicit indices will be used, the field(s) $\phi$ are supposed to have a spatial dependence. 
The partition function for this Hamiltonian is given by:
\begin{equation}
Z=\int d\phi \sum_n \langle n | \exp(-\beta\hat{H}(\phi)+\beta\mu\hat{N}) |n\rangle
\end{equation}
where $|n\rangle$ are the eigenvectors of the one-electron sector. 
To calculate the observables, an arbitrary configuration of classical fields $\phi$ is selected as a starting point.
The Boltzmann weight or action of that configuration, $S(\phi)$, is calculated by diagonalizing the one-electron sector at
finite temperature. Then a small local change of the field configuration is proposed, so that the new configuration
is denoted by $\phi'$ and its action is recalculated to obtain the difference in action $\Delta S = S(\phi')-S(\phi)$
. This new configuration is accepted or rejected based on a
Monte Carlo algorithm like Metropolis or heat bath and the cycle starts again. 
In short, the observables are traditionally 
calculated
using exact 
diagonalization of the one-electron sector at every spin ``flip'' and Monte Carlo integration for the
classical fields \cite{re:dagotto02}. The PEM/TPEM substitutes the diagonalization for a
polynomial expansion and the details are presented in
Ref.~\cite{re:motome99,re:furukawa01,re:furukawa03}.\\
It will be assumed that the Hamiltonian $H(\phi)$ is normalized, which simply implies a re-scaling:
\begin{eqnarray}
H&\rightarrow& (H-b)/a\nonumber\\
a&=&(E_{max}-E_{min})/2\nonumber\\
b&=&(E_{max}+E_{min})/2,\label{eq:hamnormalization}
\end{eqnarray}
where $E_{max}$ and $E_{min}$ are the maximum and minimum eigenvalues of the original Hamiltonian respectively.
This in turn implies that the normalized Hamiltonian has eigenvalues $\epsilon_v\in[-1,1]$. 
The values of $E_{max}$ and $E_{min}$ depend
on the particular Hamiltonian under consideration and should be calculated in advance.\\
The observables that can be calculated directly\footnote{In principle, it would be
possible to calculate more complicated observables by expanding not only the density of
states but also off-diagonal elements, 
\eg\ $\langle{\rm\hat{T}}{\hat c}^\dagger_{i\sigma}(t) {\hat c}_{j\sigma}(0)\rangle$.} 
with the TPEM fall into two categories:
(i) those that do not depend directly on fermionic operators, \eg\ 
 the magnetization, susceptibility and classical spin-spin correlations 
 in the double
exchange model and
(ii) those for which a function $F(x)$ exists
such that they can be written as:
\begin{equation}
A(\phi)=\int_{-1}^{1}F(x)D(\phi,x)dx,
\label{eq:observable}
\end{equation}
where $D(\phi,\epsilon)=\sum_\nu\delta(\epsilon(\phi)-\epsilon_\nu)$, 
and $\epsilon_\nu$ are the eigenvalues of $H(\phi)$, \ie\ $D(\phi,x)$ is the density of
states of the system. 
For the former, the calculation is straightforward and simply involves the average over 
Monte Carlo configurations.
For the latter, a function $F(x)$ can be expanded in terms of Chebyshev
polynomials in the following way:
\begin{equation}
F(x)=\sum_{m=0}^{\infty}f_mT_m(x),
\label{eq:coeffs}
\end{equation}
where $T_m(x)$ is the $m-$th Chebyshev polynomial evaluated at $x$.
Let $\alpha_m=2-\delta_{m,0}$. The coefficients $f_m$ are calculated using the formula:
\begin{equation}
f_m=\int_{-1}^1 \alpha_m F(x)T_m(x)/(\pi\sqrt{1-x^2}).
\label{eq:coeffscalculation}
\end{equation}
The moments of the density of states are defined by:
\begin{equation}
\mu_m(\phi)=\sum_{\nu=1}^{N_{dim}}\langle \nu| T_m(H(\phi))|\nu\rangle,
\label{eq:moments}
\end{equation}
where $N_{dim}$ is the dimension of the one-electron sector.
Then, the observable corresponding to the function $F(x)$, can be calculated using:
\begin{equation}
A(\phi)=\sum_m f_m \mu_m(\phi)
\label{eq:expansion}
\end{equation}
In practice, the sum in Eq.~(\ref{eq:expansion}) is performed up 
to a certain cutoff value
$M$, without appreciable loss in accuracy as described in Ref.~\cite{re:motome99,re:furukawa01}.
The calculation of $\mu_m$ is carried out recursively.
$|\nu;m\rangle  =  T_m(H)|\nu\rangle=2H|\nu;m-1\rangle-|\nu;m-2\rangle$
and hence:
\begin{eqnarray}
\mu_{2m} &=&  \sum_\nu(\langle m;\nu|\nu;m\rangle -1 )\nonumber\\
\mu_{2m+1} &=& \sum_\nu(\langle m;\nu |\nu;m\rangle +1 -\langle\nu;0|\nu;1\rangle),
\end{eqnarray}
are used to calculate the moments.
The process involves a sparse matrix-vector product, \eg\ in $T_m(H)|\nu\rangle$,
  yielding a cost  of
$O(N^2)$ for each configuration, \ie\ $O(N^3)$ for each Monte Carlo step. In addition,
an improvement of the present method has been proposed \cite{re:furukawa03} based on 
a truncation of the matrix-vector product mentioned before
and it turns out that the resulting algorithm has a complexity linear in $N$. This approximation
is controlled by the small parameter $\epsilon_{pr}$.\\
Moreover, for the Monte Carlo integration procedure, the difference in action,
$\Delta S = S(\phi')-S(\phi)$  has to be computed at
every step. Since this operation requires calculating two set of moments, for
$\phi$ and $\phi'$, the authors of Ref.~\cite{re:furukawa03} have also developed a truncation procedure for this
trace operation controlled by a small parameter, $\epsilon_{tr}$.
This truncation is based on the observation that if $\phi$ and $\phi'$ differ only in very few sites then the
corresponding moments will differ only for certain indices allowing for a significant reduction
of the computational complexity. 
%Given two matrices $H(\phi)$ and $H(\phi')$ with respective eigenvalues
%$\epsilon_v(\phi)$ and $\epsilon'_v(\phi)$, the difference in moments between the two is
%simply given by:
%\begin{equation}
%\mu_m-\mu'_m=\sum_{\nu=1}^{N_{dim}}T_m(\epsilon_\nu(\phi))-T_m(\epsilon_\nu(\phi')).
%\label{eq:diff}
%\end{equation}
The TPEM library presented here 
implements this truncation as well.\\
In what follows, the size of the Hilbert space will be denoted by $N_{dim}$ and it will depend on the size
 of the lattice as well as on the particular model to be studied. For a one-band 
double exchange
model on a lattice of $N$ sites and finite coupling, $N_{dim}=2N$; the factor of 2 accounts for the spin degree of freedom. 
\section{The Library}\label{sec:implementation}
\subsection{Implementation}

The code is written in C and can be called from a C or C++ program. If the library is
compiled statically the file libtpem.a is produced. To use the functions provided by the
library the header file ``tpem.h" has to be included.\\
In the code, complex numbers are simply represented by the structure:\\
\begin{tt}
typedef	struct \{ double real, imag; \} tpem\_t;\\
\end{tt}
As mentioned before, matrix-vector multiplications must be performed in a sparse way, 
\ie, multiplications
that yield a null result must be avoided for efficiency.
The structure tpem\_sparse, defined in tpem\_sparse.c,
 implements a sparse matrix in compressed row storage (CRS)
format.
The CRS format puts the subsequent nonzero elements of the matrix rows
 in contiguous memory locations. We create 3 vectors: one for complex numbers containing the values of the
 matrix entries 
   and the other two for integers ($colind$ and $rowptr$).
  The vector $values$ stores the values of the non-zero elements of the matrix,
   as they are traversed in a row-wise fashion.
  The $colind$ vector stores the column indices of the elements of the $values$
  vector. That is, if
 	$values[k] = a[i][j]$ then $colind[k] = j$.
  The $rowptr$ vector stores the locations in the $values$ vector that start
  a row, that is
 	$values[k] = a[i][j]$ if $rowptr[i] \le i < rowptr[i + 1]$.
  By convention, we define $rowptr[N_{dim}]$ to be equal to the number of non-zero elements,
   $n_z$, in the
  matrix. The storage savings of this approach are significant since instead of
  storing $N_{dim}^2$ elements, we need only $2n_z + N_{dim} + 1$ storage locations.\\
  To illustrate how the CRS format works, consider the non-symmetric matrix defined by
  \begin{equation}
  A=\left[\begin{tabular}{llllll}
  
 	10 &  0 & 0 & 0  & -2 & 0 \\
 3 &  9 &  0 &  0 &  0 &  3 \\
	 0 &  7 &  8 &  7 &  0 &  0 \\
 3 &  0 &  8 &  7  & 5 &  0 \\
 0 &   8 &  0 &  9 &  9 & 13 \\
 0 &  4 &  0 &  0 &  2&  -1 \\
\end{tabular}\right]\end{equation}
 The CRS format for this matrix is then specified by the arrays:\\
 \begin{tt}
  values = [10 -2  3  9  3  7  8  7  3 ... 9 13  4  2 -1 ]\\
 colind = [ 0  4  0  1  5  1  2  3  0 ... 4  5  1  4  5 ]\\
 rowptr = [ 0  2  5  8 12 16 19 ]\\
\end{tt}
Besides the obvious saving in storage, CRS format allows for a model independent library implementation
and an easy algorithm for matrix-vector multiplication as shown in
Fig.~\ref{fig:matrixvector}.
\begin{figure}
\begin{tt}
\begin{Verbatim}[frame=single,framesep=5mm]
void tpem_sparse_mult (sparse *matrix, tpem_t *dest, 
tpem_t *src)
{
  size_t row, col, k;
  tpem_t tmp;
 
  for (row = 0; row < matrix->rank; row++) {
    sum = 0.0;
    for (k=matrix->rowptr[row];k<matrix->rowptr[row+1];k++) {
      col  = matrix->colind[k];
      sum += matrix->values[k] * src[col];
    }
    dest[row] = sum;
  }
}
\end{Verbatim}
\end{tt}
\caption{Matrix-vector multiplication function using the CRS format. 
$src$ contains the vector to be multiplied and the results are stored in $dest$.
\label{fig:matrixvector}}
\end{figure}

The truncations in the matrix-vector product and in the  action difference 
are calculated with the aid of tpem\_subspace.c which implements a simple stack. The
stack is used to hold a ``subspace'' of kets of the one-electron Hilbert space that grows dynamically. It
is in this subspace that matrix-vector multiplications are performed instead of using the complete Hilbert
space.

\subsection{Functions provided by the Library}
\begin{tt}
\noindent {\bf 1.} void tpem\_init();
\end{tt}

\noindent Description: It must be called before using the library.

\begin{tt}
\noindent {\bf 2.} void tpem\_calculate\_coeffs
(size\_t M, double *coeffs,\\
double (*G)(int,double));
\end{tt}

\noindent Description: It calculates $f_m$ using Eq.~(\ref{eq:coeffscalculation}).

\noindent Arguments:

\begin{itemize}
\item {\tt M}: cutoff (input)
\item {\tt coeffs}: array of doubles where the coefficients $f_m$, Eq.~(\ref{eq:coeffs}) will be
stored. (output)
\item {\tt double 
(*F)(double)}:  The function corresponding to the observable that we want to expand as given by
Eq.~(\ref{eq:observable}). The function takes a {\tt double} and returns a {\tt double} (input).
\end{itemize}

\begin{tt}
\noindent {\bf 3.} void tpem\_calculate\_moment\_tpem
(tpem\_sparse *matrix, size\_t M, double *moment,double epsProd);
\end{tt}

\noindent Description: It calculates the moments of the density of states, $\mu_m$, 
as defined by Eq.~(\ref{eq:moments}). The method used is TPEM as described in
\cite{re:furukawa03}.

\noindent Arguments: 

\begin{itemize}
\item {\tt matrix}: the matrix in compressed row storage (input).
\item {\tt M}: the cutoff (input).
\item {\tt moment}: array of {\tt doubles} to store the moments, Eq.~(\ref{eq:moments})
(output).
\item {\tt epsProd}: the tolerance for the matrix-vector product truncation (input).
\end{itemize}

\begin{tt}
\noindent \bf{4.} void tpem\_calculate\_moment\_pem
(tpem\_sparse *matrix, size\_t M, double *moment);
\end{tt}

Description: Same as previous but it uses PEM algorithm.

\begin{tt}
\noindent void	tpem\_calculate\_moment\_diff\_tpem
	(tpem\_sparse *matrix0, tpem\_sparse *matrix1,
					size\_t M, double *moments,
					size\_t n\_support,		
					size\_t *support, double epsTrace, double epsProd);
\end{tt}

\noindent Description: It calculates the difference in moments for two matrices, 
using the trace
truncation algorithm. 

\noindent Arguments:

\begin{itemize}
\item {\tt matrix0}: The first matrix (input).
\item {\tt matrix1}: The second matrix (input).
\item {\tt M}: the cutoff (input).
\item {\tt moments}: array of {\tt doubles} to store the difference in moments (output).
\item {\tt n\_support}: Number of entries where the two matrices differ (input).
\item {\tt support}: Vector containing the column index of the entries where the two matrices differ. 
For example, in the double exchange model with finite coupling if site $i$ is being updated then
{\tt support=[i,i+N]} where {\tt N} is the number of sites (input).
\item {\tt epsTrace}: The tolerance for the trace truncation algorithm (input).
\item {\tt epsProd}: The tolerance for the matrix-vector multiplication truncation (input). 
\end{itemize}
Note that this function does not calculate the moment difference for any two matrices, only for matrices that differ
in indices specified by {\tt n\_support} and {\tt support} as explained above.

\begin{tt}
\noindent \bf{5.} void	tpem\_calculate\_moment\_diff\_pem
	(tpem\_sparse *matrix0, tpem\_sparse *matrix1,
					size\_t M, double *moments);
\end{tt}

\noindent Description: It calculates the difference in moments for two matrices
 without approximations. This function is provided for easy
integration of PEM and TPEM algorithms. Since there is no truncation, the support array,
its size and the tolerances are not needed.

\begin{tt}
\bf{6.} double tpem\_expansion (size\_t M, double *moments,
double *coeffs)
\end{tt} 	

Description: Given $f_m$ and $\mu_m$ calculates $A(\phi)$ as given by
Eq.~(\ref{eq:expansion}).

\begin{tt}
\noindent {\bf 7.} void tpem\_finalize();
\end{tt}

\noindent Description: It can be called to free the resources used by the library and reset all input.
 
\section{Simple Examples}\label{sec:simpleexamples}

\subsection{Calculating an Integral}

\begin{figure}
\centering{
\includegraphics[width=6.5cm]{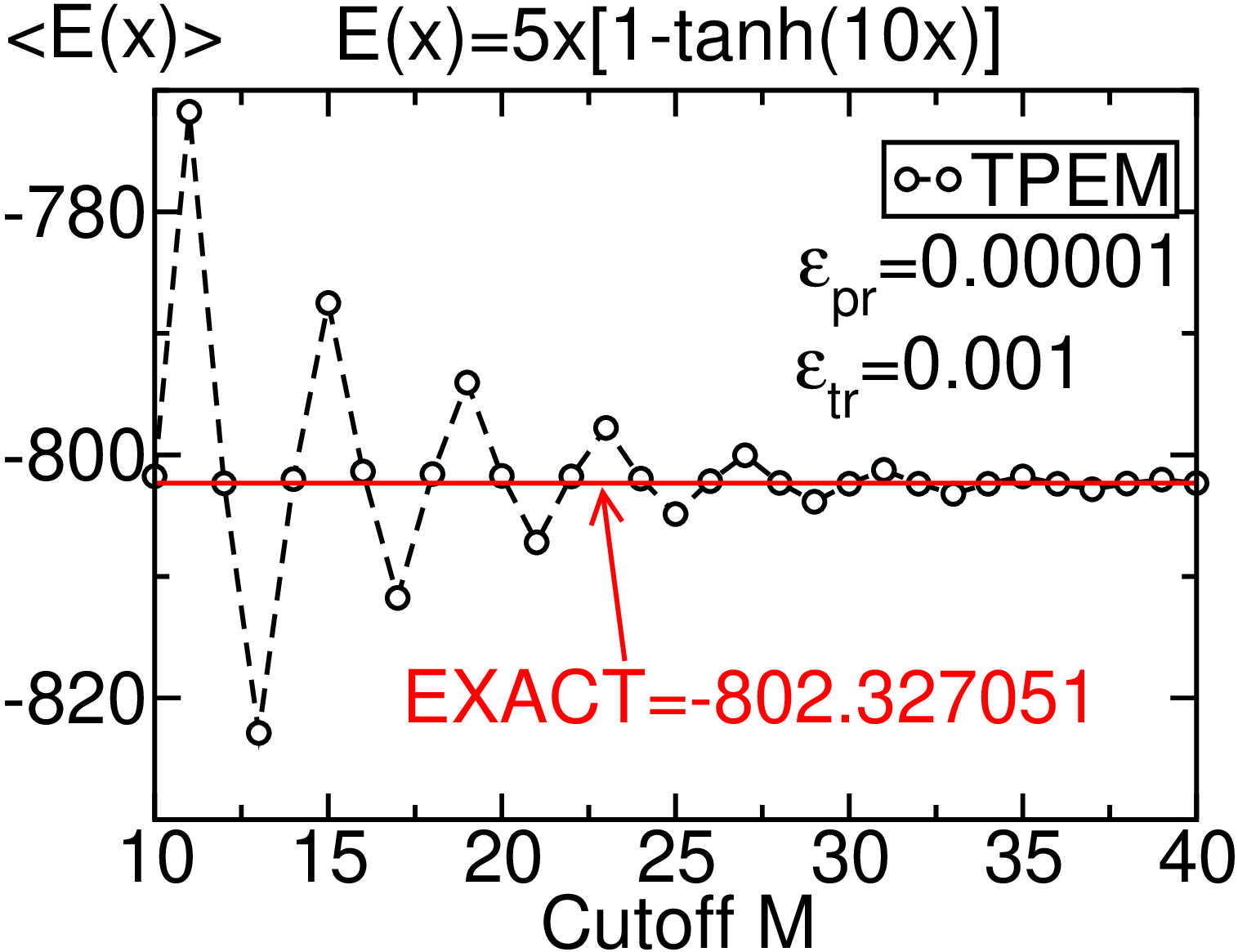}
\includegraphics[width=6.5cm]{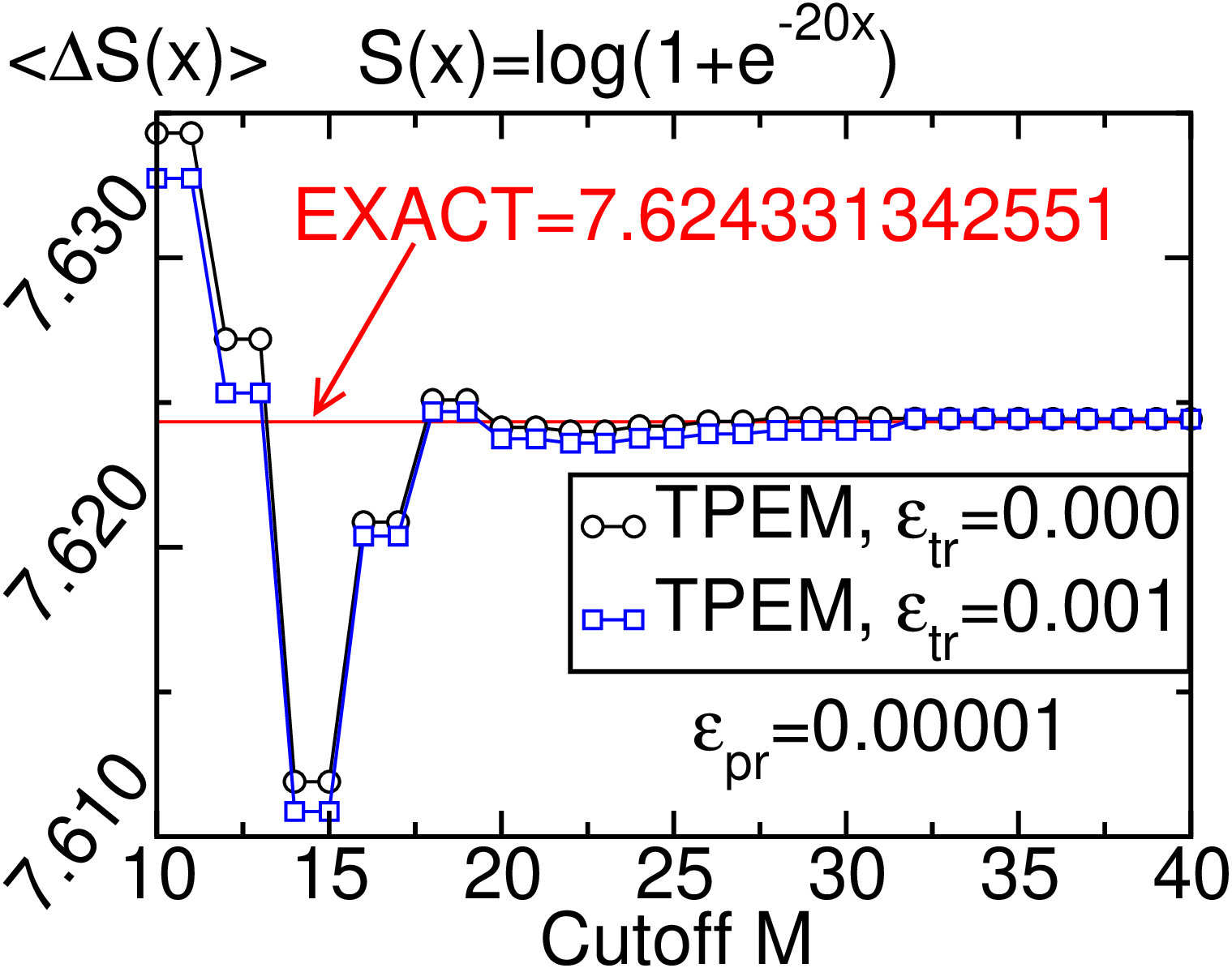}
}
\caption{(a) $\langle E(x)\rangle$ calculated using the TPEM with the parameters shown vs. the cutoff,
$M$. The exact value is also indicated. (b) $\langle\Delta S(x)\rangle$ calculated using the 
TPEM with and without trace truncation vs. the cutoff,
$M$. In the former case $\epsilon_{tr}=0.001$ and in the latter $\epsilon_{tr}=0$. 
The exact value is also indicated.\label{fig:sampleenergy}}
\end{figure}

To illustrate the use of the library several integrals will be calculated based on a
density of states, $D(x)$, given by the one-dimensional 
spinless double exchange model with random potentials, $V_i$, whose Hamiltonian is:
\begin{equation}
{\hat H}=-t\sum_{<ij>,\sigma}{\hat c}^\dagger_{i\sigma} {\hat c}_{j\sigma} +\sum_i V_i\hat{n}_i.
\label{eq:hamspinless}
\end{equation}
The complete code discussed in this section is provided in the file tpem\_test.c. The most important steps will be
explained here.
The matrix used is $400\times400$, is calculated in the CRS format and normalized so that its eigenvalues are in the
interval $[-1,1]$ as explained in Section~\ref{sec:overview}  and in Eq.~(\ref{eq:hamnormalization}). 
The library must be initialized by calling the function {\tt tpem\_init}. 
Let us define $E(x) = 5.0 x (1.0 - \tanh (10.0 x))$\footnote{This function emulates
an energy function for a fermionic system.} and calculate 
$\int_{-1}^{1}D(x)E(x)dx$ applying both exact diagonalization and the TPEM.
In tpem\_test.c the diagonalization is done by calling the function
{\tt diag\_apply}. Next the integral is performed using the TPEM for different values of the cutoff, $M$, and fixed
$\epsilon_{pr}=10^{-5}$ and $\epsilon_{tr}=10^{-3}$ in the function {\tt tpem\_apply}.
The code is self explanatory and shows the ease of use of the library:
First, the coefficients need to be calculated using:\\
\begin{tt}
tpem\_calculate\_coeffs (cutoff, coeffs, funcptr);\\
\end{tt}
Next, the moments are obtained by calling:\\
\begin{tt}
tpem\_calculate\_moment\_tpem(matrix, cutoff, moment,eps);\\
\end{tt}
Finally, the integral is calculated simply by multiplying the moments times the coefficients:
\begin{tt}
ret = tpem\_expansion (cutoff, moment, coeffs);\\
\end{tt}
The output of the program is presented at the end and in Fig.~\ref{fig:sampleenergy}a.
Similarly, other integrals are calculated in tpem\_test.c with the function 
$N(x) =  0.5  (1.0 - \tanh (10.0  x))$. In both cases, it can be seen that after $M\approx 30$ the results agree with
the ones obtained by applying the traditional diagonalization method. Moreover, if only 
$M$ even is considered then the convergence for $E(x)$ is achieved for a much smaller
value of $M$, namely $M\approx10$. 

\subsection{Using Trace Truncation}

The last part of tpem\_test.c tests the trace truncation. Consider two matrices corresponding
to a one-dimensional 
spinless double exchange model with random potentials, Eq.~(\ref{eq:hamspinless}), 
that differ only in the value
of the potential at the first site.
Consider the function, $S(x)=\log (1.0 + \exp (-10.0 x))$, which emulates the action. The testing
program calculates the difference in $S(x)$ for both matrices in three ways: (i) with the exact
diagonalization, (ii) by using the TPEM without trace truncation and (iii) by using trace truncation.
The last two results are parameterized in terms of $M$ and both assume $\epsilon_{pr}=10^{-5}$ whereas
$\epsilon_{tr}=0$ when no trace truncation is used and $\epsilon_{tr}=10^{-3}$ in the second case.\\
The results are presented in Fig.~\ref{fig:sampleenergy}b. Both TPEM calculations agree with the exact diagonalization after
$M\approx20$.\\
\begin{figure}
\centering{
\includegraphics[width=6.5cm,clip]{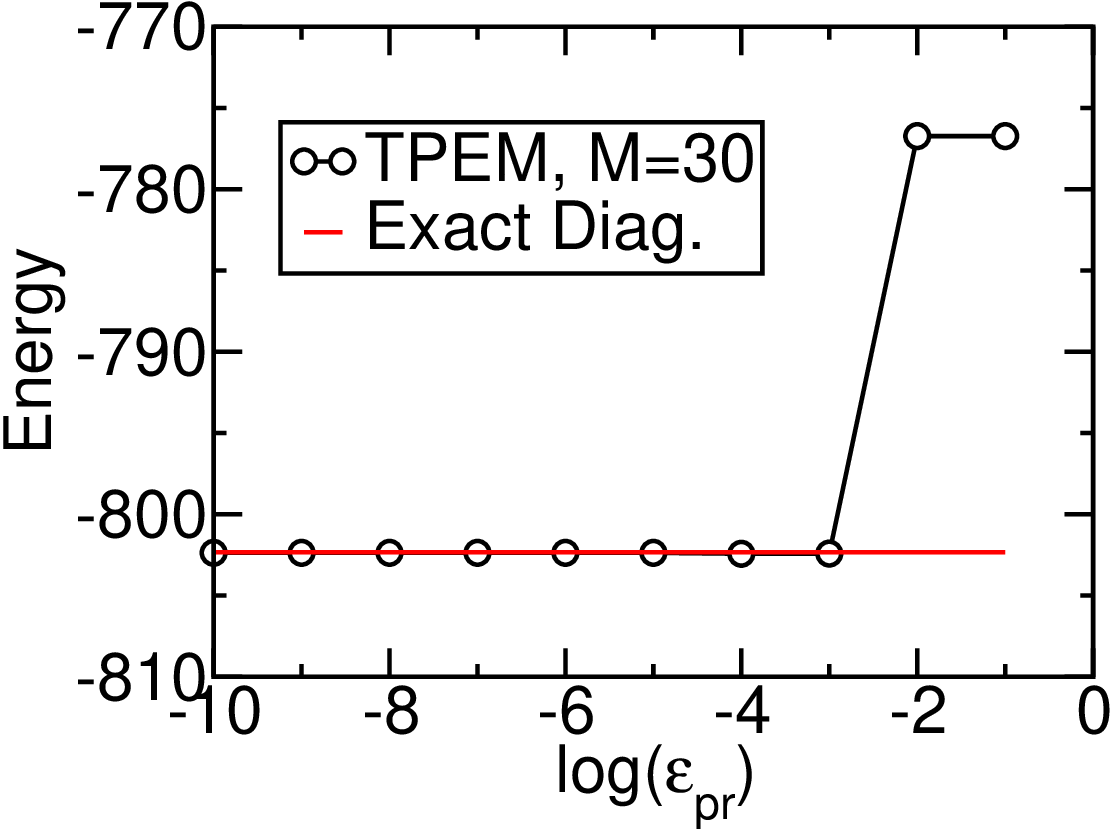}
}
\caption{TPEM calculated $\langle E\rangle$ for different values of $\epsilon_{pr}$. The
$x-$axis is given in logarithmic scale. The value of $\langle E\rangle$ from the exact
diagonalization technique is also indicated.  \label{fig:evsepspr}}
\end{figure}
The dependence of the quality of the results on $\epsilon_{pr}$ is shown in
Fig.~\ref{fig:evsepspr} for the function $E(x)$ where it can be seen that
 $\epsilon_{pr}=10^{-3}$ is enough to obtain
very high accuracy for this model. However, for the systems that will be described in the
next section we have found that $\epsilon_{pr}$ should be as small as $10^{-5}$ for the
results to be accurate, and this value will be used in the rest of this work.

\section{Advanced Applications: TPEM and Monte Carlo}\label{sec:applications}

\subsection{Double Exchange Model at Finite Coupling}
\begin{figure}
\centering{
\includegraphics[width=10cm,clip]{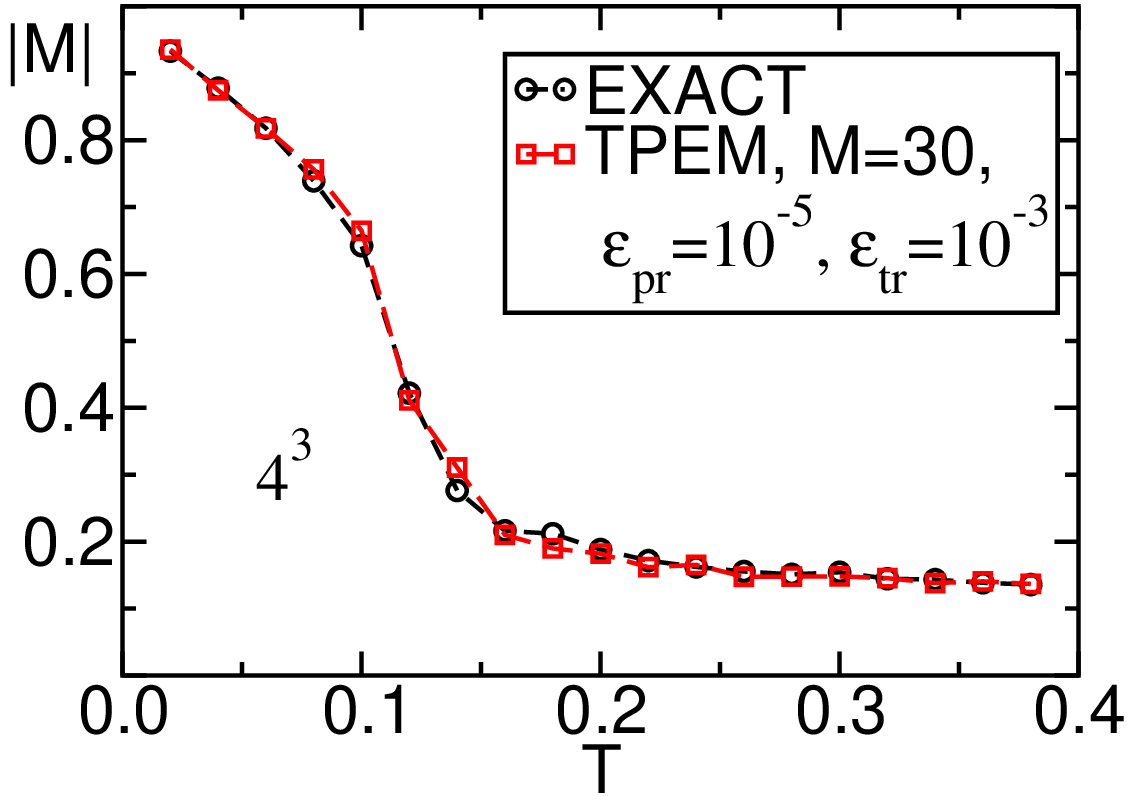}
}
\caption{Magnetization vs. temperature, $T$, on a $4^3$ lattice for model
Eq.~(\ref{eq:hammanganites}), calculated using both exact diagonalization and the TPEM
with the indicated parameters. The maximum possible value of $|M|$ is 1.\label{fig:magnetization4d3}}
\end{figure}

\begin{figure}
\centering{
\includegraphics[width=6.5cm,clip]{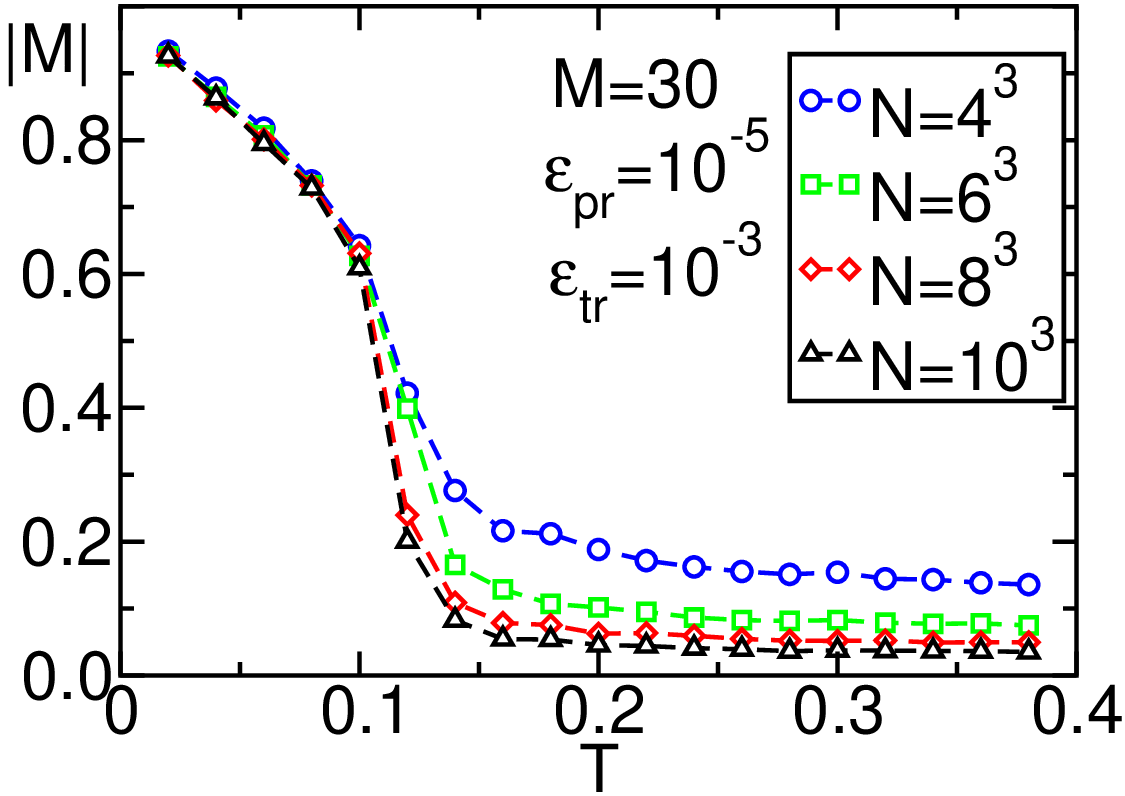}
\includegraphics[width=6.5cm,clip]{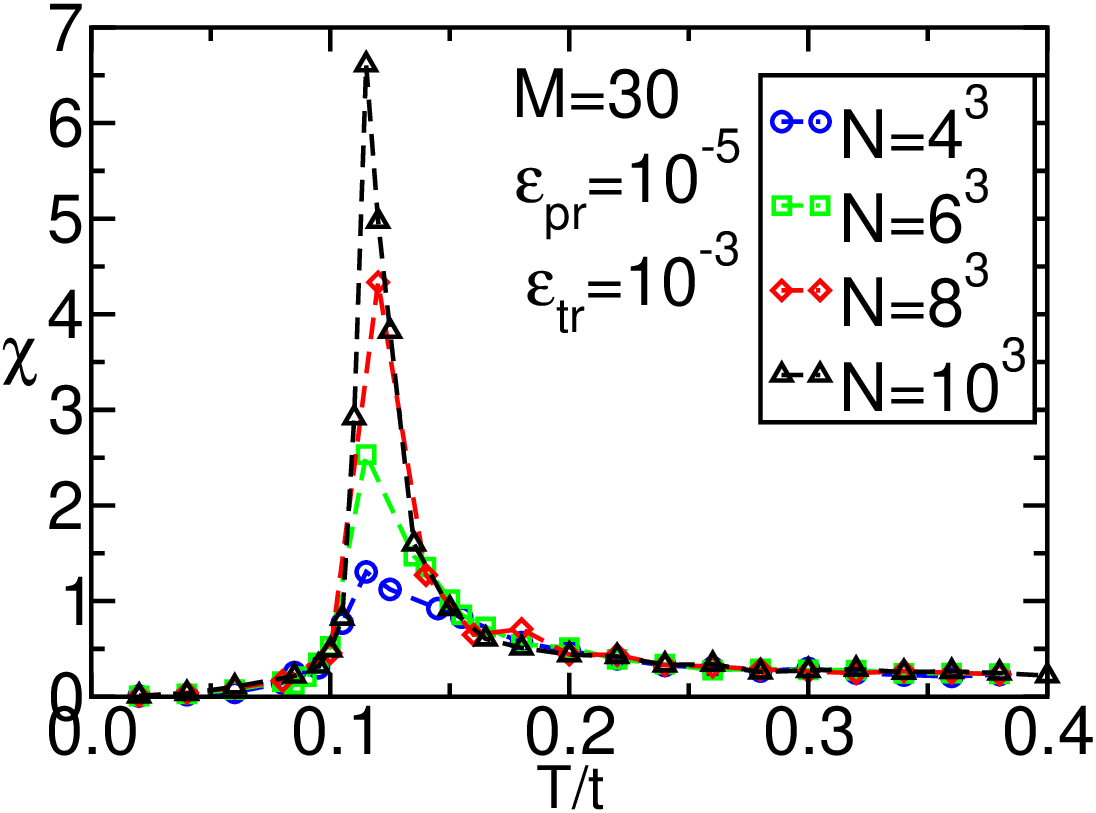}
}
\caption{TPEM calculated (a) Magnetization vs. $T$  and (b)
$\chi$ vs. $T$ for the lattices and parameters
indicated.\label{fig:magAndChi}}
\end{figure}

Double exchange models appear in the description of the 
colossal magnetoresistance effect (CMR) in manganites where the electron-phonon coupling and the 
Coulomb interactions are usually
neglected\cite{re:dagotto01}. These models can correctly produce ferromagnetic phases 
as long as the Hund coupling is large enough. In this case, the electrons directly jump
from manganese to manganese spin and their kinetic energy is minimized if these spins are
aligned.
\indent The Hamiltonian of the system in the one-band approximation can be written 
as \cite{re:zener52,re:furukawa94}:
\begin{equation}
{\hat H}=-t\sum_{<ij>,\sigma}{\hat c}^\dagger_{i\sigma} {\hat c}_{j\sigma} -
J\sum\vec{S}_{i}\cdot\vec{\sigma}_i,
\label{eq:hammanganites}
\end{equation}
\noindent where 
${\hat c}^\dagger_{i\sigma}$ creates a carrier at site $i$
with spin $\sigma$. The carrier-spin operator interacting
ferromagnetically with the localized Mn-spin $\vec{S}_i$ is
$\vec{\sigma}_i=\sum_{\alpha,\beta}{\hat
c}^\dagger_{i\alpha}\vec{\sigma}_{\alpha,\beta}{\hat c}_{i\beta}$. 
On a cubic lattice of 
dimension $D$ the largest and smallest spectrum bounds of Hamiltonian Eq.~(\ref{eq:hammanganites}) are $E_{min}=-2tD-J$ and 
$E_{max}=-2tD+J$, respectively. 
To test the TPEM for this physical model, we start with the interesting case of a
ferromagnetic to paramagnetic transition. The coupling $J$ is chosen to be $J=8t$ and
the electronic density is adjusted with $\mu=-8t$ to obtain a quarter filling, \ie\
$n=0.5$. We have performed 1,000 thermalization and 1,000 measurements which were enough
to achieve both convergence and small errors. The results for the magnetization of the
system,
defined as 
\begin{equation}
|M|=\frac 1N|\sum_i\vec{S}_i|
\label{eq:magnetization}
\end{equation} 
are shown in Fig.~\ref{fig:magnetization4d3} and compared to the traditional 
exact diagonalization calculation,
where the high accuracy of the method can be seen clearly.
We repeat the calculations for larger lattices and also measure the magnetic
susceptibility, $\chi$, as a function of the temperature (see Fig.~\ref{fig:magAndChi}).\\ 
The boundary conditions used are anti-periodic in one direction and periodic in the other
two. This is a numerical trick in the sense that it is an effective way to lift the degeneracy due to
small size lattices. This degeneracy affects the form of the density of states making it difficult to expand it
when performing simulations on $4^3$. The effect is less and less relevant as size increases. Moreover, the choice of
boundary conditions does not matter in the thermodynamic limit. Twisted boundary conditions has been used
extensively in numerical simulations \cite{re:assaad02,re:nakano99}.

\begin{figure}
\centering{
\includegraphics[width=6.5cm,clip]{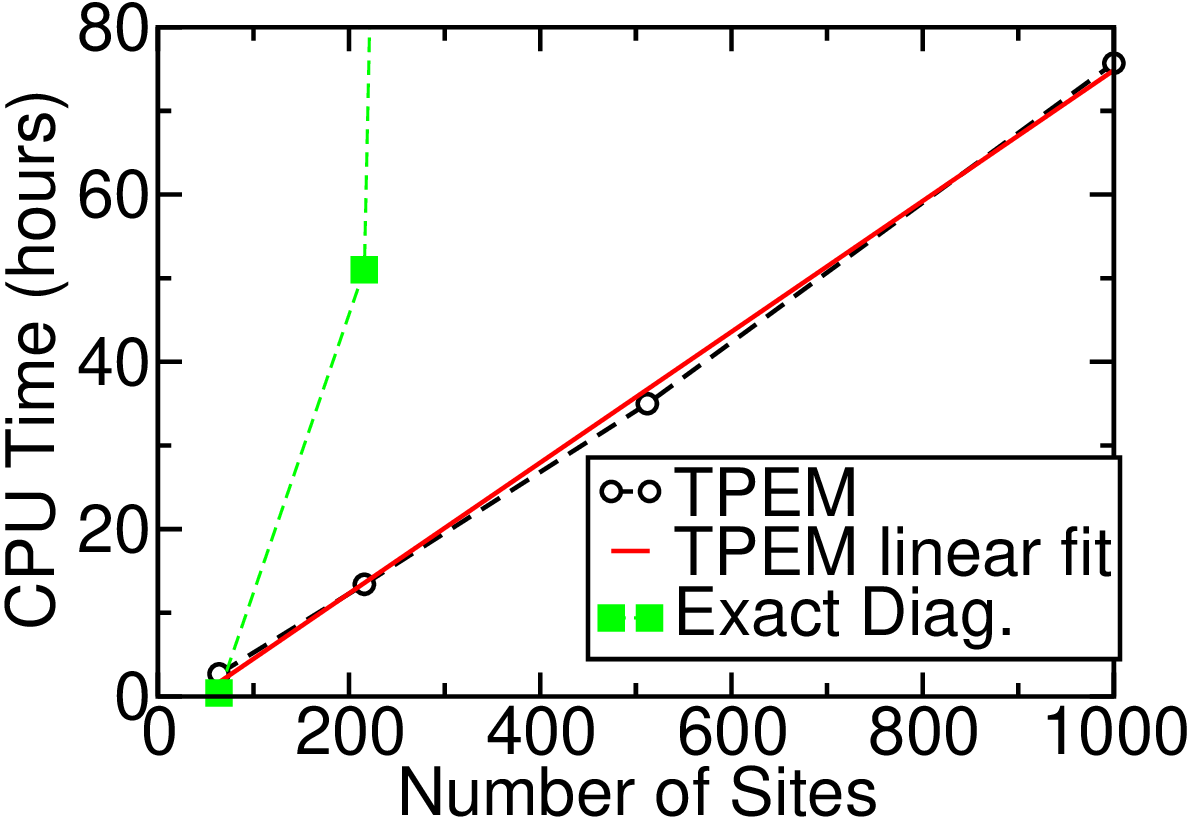}
\includegraphics[width=6.5cm,clip]{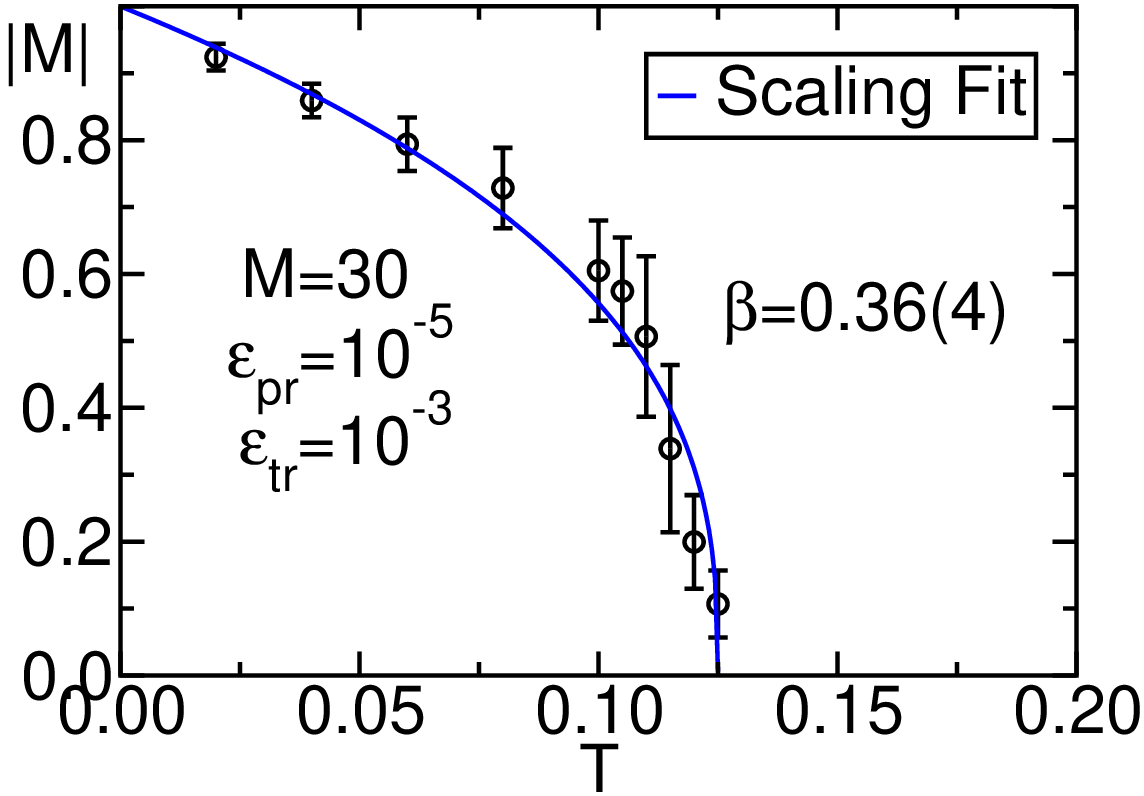}
}
\caption{(a) CPU Times for the TPEM algorithm applied to model 
Eq.~(\ref{eq:hammanganites}) using 2,000 Monte Carlo steps, $M=30$, $\epsilon_{pr}=10^{-5}$, 
$\epsilon_{tr}=10^{-3}$ on a 3.06GHz PC. The linear fit yields
$y=0.0782x-3.3271$. The CPU times required for the exact diagonalization procedure are
also shown. (b) $|M|$ vs. $T$ for size-extrapolated data (lattices used: $4^3$, $6^3$, $8^3$ and $10^3$)
 and estimation of the critical exponent $\beta$.\label{fig:scale}}
\end{figure}

The CPU times to perform these
computations are shown in Fig.~\ref{fig:scale}a making use of a conventional cluster of linux PCs
with 3.06GHz of clock frequency each. Even using commodity PCs the CPU user time to perform
calculations on the largest cluster studied, $10^3$, was less than 3 days. 
The results show that the CPU time scales linearly with the size of the system as predicted by
the theory \cite{re:furukawa03}.  Moreover, the algorithm can be parallelized.
This is because calculation of the moments in Eq.~(\ref{eq:moments}) is 
completely independent for each basis 
ket $|\nu\rangle$. In this way the basis can be partitioned in such a way that each processor
calculates the moments corresponding to a portion of the basis. The CPU time to calculate the
moments on each processor is proportional to $N_{dim}/N_{PE}$, where $N_{PE}$ is the
number of processors. It is important to remark that the data to be moved between
different nodes are small compared to calculations in each node, the communication time is
proportional to $MN_{PE}$. Communication among nodes is mainly done here to add up all the
moments. A version of the TPEM library that supports parallelization will be available in the future.

The value of $T_{\rm C}$ obtained from
the $\chi$ vs. $T$ curves is approximately $T_{\rm C}=0.12t$ at $J=8t$ in very good agreement with previously
calculated values \cite{re:motome00,re:yunoki98}. In addition, we calculated the scaling coefficient
$\beta$ defined by $|M|\propto(T_C-T)^\beta$, after having made a
 size-extrapolation, \ie, after taking the thermodynamic limit. 
The result is shown in Fig.~\ref{fig:scale}b
and is within the error margin of the value given for the Universality class of the Heisenberg model. 
More information about the determination of critical exponents can be found in
Ref.~\cite{re:pelissetto02}.

\subsection{Diluted Systems}
\begin{figure}
\centering{
\includegraphics[width=8cm,clip]{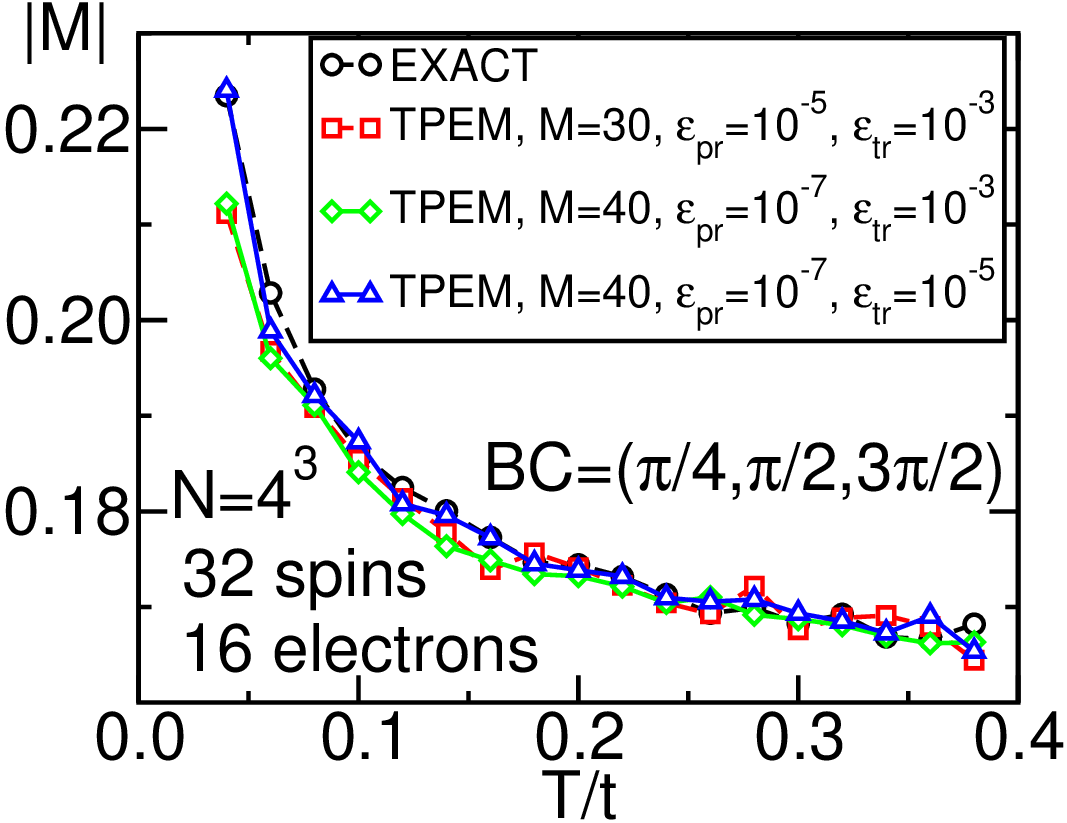}
}
\caption{$|M|$ vs. $T/t$ on a $4^3$ lattice calculated using the TPEM with the parameters
indicated.
The result given by the diagonalization method is also shown.
10,000 thermalization Monte Carlo iterations and 10,000 measurements were performed. The same random configuration
for the spatial location of the classical spins was considered in all cases.\label{fig:diluted}}
\end{figure}
The Hamiltonian for the diluted spin-fermion model that will be considered here 
is given by Eq.~(\ref{eq:hammanganites}) except that the exchange
term is replaced by $J\sum_{i\in I}\vec{S}_{i}\cdot\vec{\sigma}_i$, \ie\ 
localized spins are
present on only a subset $I$ of the lattice sites. 
Through nearest-neighbor hopping, 
the carriers can hop to \emph{any} site of the square or cubic lattice.
In the same way as in the case of the double-exchange model,  
 for diluted systems we have used periodic boundary conditions for faster convergence. These are specified by the
phases $(\pi/4,\pi/2,3\pi/4)$ in the $x$, $y$ and $z$ directions respectively.\\
It is important to remark that Monte Carlo calculations on diluted
spin systems with concentration $x<1$ are $1/x$ times faster than the concentrated case since there are
less sites with spins to propose a spin change.\\
In this case the density of states will have a
more complicated shape, usually including a small impurity band. Therefore, it is interesting to see
whether the TPEM is capable of treating this case. 
The comparison is provided in Fig.~\ref{fig:diluted}
for a concentration of 32
spins on a $4^3$ lattice with approximately 16 electrons, where it can be seen that
the TPEM algorithm converges for $M=40$, $\epsilon_{tr}=10^{-7}$ and $\epsilon_{pr}=10^{-5}$.
  This simple test shows that even in the
case of systems with impurity bands and positional disorder, the expansion yields results
compatible with the exact treatment. Therefore, there
 is much potential for the use of this technique in the area of diluted magnetic
 semiconductors.
\subsection{Convergence}
The expansion parameters required for convergence, \ie\ the cutoff $M$ and the thresholds $\epsilon_{tr}$ and
$\epsilon_{pr}$, can be calculated on a small lattice where the exact diagonalization technique can be
used to check the T/PEM algorithm. Since these numbers do not depend on the size of the system (only on the model,
see \cite{re:furukawa03}) then they can be safely used on larger lattices.
This is shown in Fig.~\ref{fig:diluted} where unlike for the concentrated system
in this case neither $M=30$ nor $\epsilon_{tr}=10^{-5}$ is
enough for convergence but $M=40$, $\epsilon_{tr}=10^{-7}$ and $\epsilon_{pr}=10^{-5}$ is required.
On the other hand, the double-exchange model, Eq.(\ref{eq:hammanganites}), with infinite
$J$ (not studied here but discussed in \cite{re:motome03}) converges with a cutoff smaller than $M=30$.
This is because the finite coupling system density-of-states is composed of two disconnected bands separated by approximately $2J$ and so the
spectrum extends over a wide range of energies whereas in the infinite coupling system there is a single connected band resulting in a faster
convergence.\\
Therefore, the reader and user of the TPEM library
should not assume that the values presented in the previous examples will guarantee convergence for a particular
model but should instead perform a check similar to the one presented in this section to determine the minimum 
value of the cutoff $M$ and the maximum values of $\epsilon_{tr}$ and $\epsilon_{pr}$ required for convergence.

\section{Conclusions}
In summary, we have provided a software library that implements the TPEM for fermion 
systems coupled to classical
fields. This library will allow theorists to study a variety of systems employing the 
TPEM at the same level that, for instance, the LAPACK library \cite{laug}
is being used for exact diagonalization.\\
The TPEM has an enormous potential. For example, studies of diluted magnetic semiconductors 
 that had not been possible
 before with more than one 
band will now be accessible and the results will be 
 presented elsewhere. These studies are crucial to understand the properties of  magnetic semiconductors and will
help in the search for similar compounds with higher Curie temperatures.
These high-Curie temperature compounds would in turn be useful for technological
 applications, for example in the fabrication of 
spin electronic or spintronic devices \cite{re:zutic04}. 
The possibility of studying larger systems will not merely imply a better
estimation of the physical observables but will allow for the study of more complex systems like 
transition metal oxides with realistic bands and
nanostructures.

\section{Acknowledgments}
This work was supported in part by NSF grants DMR-0122523, DMR-0312333, and DMR-0303348.
G.A. performed this research as a staff member at the Oak Ridge National Laboratory, managed by UT-Battelle, LLC, 
for the U.S. Department of Energy under Contract DE-AC05-00OR22725.
 Most calculations
were performed on the CMT computer cluster at the NHMFL and we acknowledge 
the help of T. Combs. J. Burgy helped with the design of the software library.
We would like to thank also K. Foster for proofreading the manuscript. 
\bibliographystyle{elsart-num}
\bibliography{thesis}

\section{Test Run Output}
\begin{tt}
\begin{verbatim}
********************************************************
****** TESTING TRUNCATED POLYNOMIAL EXPANSION **********
********************************************************

This testing program calculates model properties in two ways:
(i)  Using standard diagonalization
(ii) Using the truncated polynomial expansion method
All tests are done for a nearest neighbor interaction with
random (diagonal) potentials.
 
-------------------------------------------------------------
TEST 1: MEAN VALUE FOR THE FUNCTION:                         
        N(x) =  0.5 * (1.0 - tanh (10.0 * x))

** Using diagonalization <N>=195.349187
** Using TPEM <N>=(cutoff--> infinity) lim<N_cutoff> where <N_cutoff> is
cutoff	<N_cutoff>	%Error(compared to diag.)
10	194.740524890737	0.311577%
11	194.921830592607	0.218765%
12	194.265186764692	0.554904%
13	194.136543222392	0.620757%
14	194.315979389497	0.528903%
15	194.408557066075	0.481512%
16	194.679187298809	0.342975%
17	194.612068307437	0.377334%
18	195.036519662839	0.160056%
19	195.085374546781	0.135047%
20	195.361097093739	0.006097%
21	195.325459995073	0.012146%
22	195.550660927328	0.103135%
23	195.576686770534	0.116458%
24	195.568643202933	0.112340%
25	195.549624351656	0.102605%
26	195.521318469307	0.088115%
27	195.535221730693	0.095232%
28	195.475638686769	0.064731%
29	195.465473123028	0.059527%
30	195.430165380516	0.041453%
31	195.437598886027	0.045258%
32	195.398432059641	0.025209%
33	195.392996060833	0.022426%
34	195.392858690195	0.022356%
35	195.396834086245	0.024391%
36	195.380978851635	0.016274%
37	195.378071566221	0.014786%
38	195.355235557817	0.003096%
39	195.357361742974	0.004185%
40	195.345473742639	0.001901%
-------------------------------------------------------------
TEST 2: MEAN VALUE FOR THE FUNCTION:                         
        E(x) = 5.0 * x * (1.0 - tanh (10.0 * x))

** Using diagonalization <E>=-802.327051
** Using TPEM <E>=(cutoff--> infinity) lim<E_cutoff> where <E_cutoff> is
cutoff	<E_cutoff>	%Error(compared to diag.)
(OUTPUT OMITTED, SEE FIG 2a)

-------------------------------------------------------------
TEST 3: MEAN VALUE AND DIFFERENCE FOR THE FUNCTION:          
        S(x) =  log (1.0 + exp (-20.0 * x))

** Using diagonalization <S[matrix0]>= 854.249004021717
** Using diagonalization <S[matrix1]>= 846.624672679166
** Using diagonalization <S[matrix1]>-<S[matrix0]>=7.624331342551
** Using TPEM <S>=(cutoff--> infinity) lim<S_cutoff>
cutoff	Delta_S_cutoff	S_cutoff[diff]	Error (to diag.)
(OUTPUT OMITTED, SEE FIG 2b)
-------------------------------------------------------------
\end{verbatim}
\end{tt}

\end{document}